\documentclass[12pt]{article}
\usepackage{geometry}
\usepackage{a4}
\usepackage{graphicx,subfigure}
\usepackage{epsf}
\usepackage{amsmath}
\usepackage{amssymb}
\usepackage{caption}
\usepackage{epstopdf}
\usepackage{enumerate}
\usepackage{array}
\usepackage{cite}
\newcommand{\be}{\begin{equation}}
\newcommand{\ee}{\end{equation}}
 
 \newcommand{\Rmnum}[1]{\expandafter\@slowromancap\romannumeral #1@}
\newcommand{\bea}{\begin{eqnarray}}
\newcommand{\eea}{\end{eqnarray}}

\begin{document}
\def\A{{\mathbb{A}}}
\def\C{{\mathbb{C}}}
\def\R{{\mathbb{R}}}
\def\s{{\mathbb{S}}}
\def\T{{\mathbb{T}}}
\def\Z{{\mathbb{Z}}}
\def\W{{\mathbb{W}}}
\begin{titlepage}
\title{Black Hole Phase Transitions and the Chemical Potential}
\author{}
\date{
Reevu Maity, Pratim Roy, Tapobrata Sarkar
\thanks{\noindent E-mail:~ reevum, proy, tapo @iitk.ac.in}
\vskip0.4cm
{\sl Department of Physics, \\
Indian Institute of Technology,\\
Kanpur 208016, \\
India}}
\maketitle
\abstract{
\noindent
In the context of extended phase space thermodynamics and the AdS-CFT correspondence, we consider the chemical potential ($\mu$) dual to the number 
of colours ($N$) of the boundary gauge theory, in the grand canonical ensemble. By appropriately defining $\mu$ via densities 
of thermodynamic quantities, we show that it changes sign precisely at the Hawking-Page transition for AdS-Schwarzschild and RN-AdS black holes in five 
dimensions, signalling the onset of quantum effects at the transition point. Such behaviour is absent for non-rotating black holes in four dimensions. 
For Kerr-AdS black holes in four and five dimensions, our analysis points to the fact that $\mu$ can change sign in the stable black hole region, i.e above 
the Hawking-Page transition temperature, for a range of angular frequencies. We also analyse AdS black holes in five dimensional Gauss-Bonnet gravity, 
and find similar features for $\mu$ as in the Kerr-AdS case. 
}
\end{titlepage}
\section{Introduction}
The study of black hole thermodynamics has been an active field of research for the past few decades. In this context, thermodynamics of anti de Sitter (AdS) black holes 
has been extensively studied, motivated by the gauge-gravity duality. In these investigations, the black hole is considered as a bulk thermodynamic 
object, with usual thermodynamic quantities like temperature, entropy etc. associated to it. Properties of black hole thermodynamics have been well established
for various classes of black holes in diverse dimensions, in the large amount of existing literature on the topic. 

One of the main features of black hole thermodynamics, in contrast with thermodynamics of ordinary matter is that the former depends on the
ensemble in which the system is described. This has to do with the fact that black hole entropy is not extensive in the thermodynamic sense, i.e it scales
as an area rather than a volume. Related to this issue is the fact that the definition of a volume seems to be subtle in the context of black holes. 
Recent progress in this direction has been made, with the proposal that the cosmological constant $\Lambda$ be considered as the pressure in the first law 
of black hole thermodynamics, and its conjugate quantity be identified as a volume \cite{Kastor}. For details about the consequences of this modification 
to black hole thermodynamics, see, e.g. \cite{Kubiznakchemistry},\cite{DolanPdVterm}. In terms of the pressure and volume, black holes show thermodynamic 
phase transitions, which resemble  the liquid-gas phase transition, the re-entrant phase transition etc., in a variety of AdS black hole systems. 
Examples of such phase transitions (in the canonical ensemble) have been known for a long time \cite{Chamblin}, but the explicit introduction of the 
pressure makes the analogy with Van der Waals systems more transparent. 
The analysis of black hole thermodynamics including the black hole pressure via the cosmological constant has been dubbed as ``extended phase space
thermodynamics'' of black holes, and has seen a flurry of activity of late (see, for example \cite{kubiznakgalaxies} and references therein). 

In fact, we can envisage two distinct approaches towards towards extended phase space black hole thermodynamics. 
The first is, as mentioned, to relate the cosmological constant $\Lambda$ to the 
black hole pressure $P$. Specifically, the relation that one uses is $P = -\Lambda/8\pi$ \cite{kubiznakPV} and a conjugate thermodynamic volume can also be 
derived \cite{cveticgibbons}. In this approach, one usually holds the Newton's constant 
(in the dimension in which the black hole lives) fixed. The second approach is to instead consider the relation between the
cosmological constant and the AdS radius, and relate the latter, by the AdS-CFT duality, to the number of colours $N$ in the dual gauge theory. In this latter approach,
one is allowed to ``vary'' the number of colours. The first work in this direction appeared in \cite{Dolanbose} (see also \cite{Caithermogeom1},\cite{Caithermogeom2}). 
This issue might be quite subtle on the gauge theory side, as we 
are effectively describing a ``flow'' in the theory space, parametrized by  $N$. Importantly, when $N$ is treated as a variable in the theory, a higher 
dimensional Newton's constant has to be held fixed (along with the Plank length in that dimension). For example, if we consider a theory on $AdS_5\times S^5$, 
the AdS length is related to the number of colours of the dual gauge theory via the ten dimensional Newton's constant $G_{10}$ and the ten dimensional 
Planck length $l_{10}$. To vary $N$ in such a theory, we fix $G_{10} = l_{10}=1$.\footnote{The Boltzmann's constant is always set to unity in this paper,
along with $\hbar$ and the speed of light, $c$.}

From the point of view of the canonical ensemble, a varying $N$ may not be of particular interest, as it simply redefines the critical values of the charges, 
and we do not expect much interesting physics to emerge. As a concrete example, let us consider the five dimensional RN-AdS black hole, which is known 
to exhibit a liquid-gas type of phase transition up to a certain critical value of the charge $Q$ and the horizon radius $r_h$. An elementary 
calculation reveals that if we relate the AdS radius $L$ via the gauge-gravity duality to $N$, by $L = N^{1/4}$ (we will not be careful about the precise 
pre-factors which can be absorbed in $N$) the Hawking temperature of the black hole can be written as 
\begin{equation}
T =\frac{1}{2\pi  r_h}+\frac{r_h}{\pi  \sqrt{N}} -\frac{8 \pi  Q^2}{3 N r_h^5}
\label{largeN}
\end{equation}
It is then easy to check that the critical charge and horizon radius at the second order phase transition are simply rescaled : 
$Q_c= \frac{N }{12 \sqrt{5} \pi }$ and $r_{h,c} = \frac{N^{1/4}}{\sqrt{3}}$. Setting $N=1$, we can recover the standard results for the situation where the
five dimensional Newton's constant $G_5$ is set to unity (without relating the cosmological constant to the pressure). An entirely similar analysis
follows for Kerr-AdS black holes. 

However, as we have already said, from a bulk gravity point of view, one can consider black hole thermodynamics in the canonical (fixed charges) or the 
grand canonical (fixed potential) ensembles. If $N$ is promoted to the level of a variable in the theory, we can either study a theory where this is held fixed, or 
one in which it can fluctuate. It is of course difficult to envisage a situation in which the system exchanges colour degrees of freedom with a surrounding bath, 
and thus one is naturally led to a description in which the number of colours is held constant (i.e it is a dialling and not a fluctuating variable). With this in mind, 
it is interesting to look at the system described by a grand canonical ensemble with respect to the other charges (say the electric charge and the angular momentum)
whose conjugate potentials are consequently held fixed.\footnote{Throughout this work, by a slight abuse of notation, we will refer to this as the grand canonical ensemble. 
It should be kept in mind however that this is really a mixed ensemble, with the number of colours held fixed, as are the other potentials.} 

The concrete question that one can 
ask now is regarding the behaviour of the chemical potential conjugate to the number of colours in such an ensemble. It is well known that in a 
grand canonical ensemble, a charged or rotating black hole exhibits the celebrated Hawking-Page phase transition (as opposed to the liquid-gas type phase transition
in the canonical ensemble as discussed earlier), reached at a temperature $T_{HP}$ where the Gibbs free energy is lower than
a given reference background.  In this ensemble, the chemical potential $\mu$, conjugate to the number of colours is an interesting object to study, since
a vanishing chemical potential indicates the onset of quantum effects in the system as it usually indicates the breakdown of (particle) number conservation.  

In the original work that mooted the idea of a variable $N$ \cite{Dolanbose}, it was shown that for the five dimensional AdS-Schwarzschild black hole, the chemical potential
changes sign at a temperature lower than $T_{HP}$, where the black hole is essentially metastable. Our observation here is that 
in the backdrop of AdS-CFT, it might be more natural (especially in the context of variable $N$) to consider densities of thermodynamic quantities at large $N$, 
and hence compute the chemical potential $\mu$ via these. One of the main purposes of this paper is to show that interesting physics emerges
when $\mu$ is computed via such densities of thermodynamic variables (mass, entropy, charge, angular momentum), obtained by dividing these
by the volume of the space in which the dual gauge theory lives. In fact, as we will show in sequel, for the five dimensional 
AdS-Schwarzschild background, the chemical potential calculated in this way changes sign precisely at $T_{HP}$, indicating that at this transition temperature,
quantum effects might become important. We demonstrate that this fact remains valid for five dimensional RN-AdS black holes as well. 

For four dimensional AdS-Schwarzschild and for the four dimensional RN-AdS black hole, this behaviour is however absent, and $\mu$ always changes 
sign in a metastable region. Interestingly, for rotating AdS black holes, we find that for sufficiently large values of the rotation parameter (close to its maximum value), $\mu$ 
can change sign in a stable black hole region, both in four and five dimensions. 

In the remainder of this paper, we will establish these facts. Towards the end, 
we also consider AdS black holes in five dimensions, modified by a Gauss-Bonnet term. For such black holes, we show that the Gauss-Bonnet parameter
behaves qualitatively like the rotation parameter, i.e for a sufficiently large parameter values, one can expect important quantum corrections in a physical
(i.e stable) black hole region. 

\section{Charged AdS Black Holes in Four and Five Dimensions}

Let us begin with the well known example of the RN-AdS black hole. We first record the expressions for a general $d+1$ dimensional hole, and then 
specialize to the case of four and five dimensions \cite{johnsoncatastrophic}. In $d+1$ dimensions, the Einstein-Maxwell theories in AdS space can be defined via the action
\begin{equation}
S = \frac{1}{16G_{d+1}\pi}\int d^{d+1}x \sqrt{-g}\left(R - F^2 +\frac{d(d-1)}{L^2}\right)
\end{equation}
which is solved, along with an appropriate gauge potential, by the metric 
\begin{equation}
ds^2 = -f(r)dt^2 + \frac{1}{f(r)}dr^2 + r^2d\Omega_{d-1}^2
\label{ndRNAdS}
\end{equation}
where $f(r) = 1 - \frac{m}{r^{d-2}} + \frac{q^2}{r^{2d-4}} + \frac{r^2}{L^2}$. The parameters $m$ and $q$ appearing in eq.(\ref{ndRNAdS}) are related to the
ADM mass $M$ and electric charge $Q$ of the black hole by 
\begin{equation}
m = \frac{16\pi G_{d+1}M}{(d-1)\omega_{d-1}};~~~q = \frac{8\pi G_{d+1} Q}{\sqrt{2(d-1)(d-2)}\omega_{d-1}}
\label{ndmq}
\end{equation}
where $\omega_{d-1} = 2\pi^{d/2}/\Gamma(d/2)$ is the volume of the unit $d-1$ sphere. 

We will first specialize to four dimensions, 
where $f(r) = 1 - \frac{m}{r} + \frac{q^2}{r^2} + \frac{r^2}{L^2}$. We remind the reader that we are working in natural units, 
$\hbar = c = 1$, and will also set the Boltzmann's constant to unity. Doing this, and setting the area of the horizon $A = 4G_4S$, where $S$ is the entropy of 
the black hole and $G_4$ is the four dimensional Newton's constant, the Smarr formula for the mass of the black hole reads \cite{CCK}
\begin{equation}
M =\frac{G_4 \left(\pi ^2 L^2 Q^2+S^2\right)+\pi  L^2 S}{2 \pi ^{3/2} L^2\sqrt{G_4 S}}
\end{equation}
We will work in the grand canonical ensemble, and hence will express all the thermodynamic variables in terms of the electric potential and the temperature. 
To this end, noting that the electric charge of the black hole $Q$ is related to its potential $\Phi$ as $Q =\frac{\Phi  \sqrt{S}}{\sqrt{G_4\pi }}$, we obtain 
the Hawking temperature of the black hole as
\begin{equation}
T = \frac{3 G_4 S+L^2 \pi\left(1 - \Phi ^2\right)}{4 \pi ^{3/2} L^2\sqrt{G_4 S}}
\label{tgc}
\end{equation}
We use eq.(\ref{tgc}) to solve for the entropy in terms of the temperature, and obtain two solutions 
\begin{equation}
S_{\pm} = \frac{\pi  L^2}{9G_4} \left(-4 \pi  L T \sqrt{4 \pi ^2 L^2 T^2+3 \Phi ^2-3}+8\pi ^2 L^2 T^2+3 (\Phi ^2-1)\right)
\label{entf}
\end{equation}
These solutions define two distinct phases, with the positive sign corresponding to the large black hole phase and the negative sign to the small
black hole phase. The small black hole phase is always unstable in the grand canonical ensemble, as can be checked. 

Now we will use the AdS-CFT dictionary to relate the four dimensional Newton's constant to the eleven dimensional one via the AdS radius, and the eleven 
dimensional Planck length, $l_{11}$. Specifically, we use $G_4 = G_{11}L^{-7}$, $L^6 = G_{11}N/l_{11}^3$ and also use the fact that $G_{11} = l_{11}^9$.\footnote{Note
that we have ignored some numerical factors in these definitions. These do not qualitatively affect the results, but their inclusion will make the expressions clumsy.} 
Doing these substitutions, the mass of the black hole simplifies to 
\begin{equation}
M = \frac{\sqrt{S} \left(\pi  N^{3/2} \left(\Phi ^2+1\right)+S\right)}{2\pi ^{3/2} N^{11/12} l_{11}}
\label{massf}
\end{equation}
We also record the expression for the temperature and the Gibbs free energy in terms $N$ 
\begin{equation}
T = \frac{N^{3/2} \pi\left(1 -\Phi ^2\right)+3 S}{4 \pi ^{3/2}N^{11/12} \sqrt{S} l_{11}},~~~{\mathcal G} = 
\frac{\sqrt{S} \left(\pi  N^{3/2} \left(1-\Phi ^2\right)-S\right)}{4\pi ^{3/2}  N^{11/12}l_{11}}
\label{tempg}
\end{equation}
We will now set the eleven dimensional Planck length, $l_{11}$ to unity. With this, we discuss the chemical potential conjugate to the number of colours. 

First we discuss the case where we use the thermodynamic variables (and not densities) to compute the same. 
In four dimensions, the number of degrees of freedom scales as $N^{3/2}$, and hence it is natural to define the chemical potential via the first law of 
thermodynamics as
\begin{equation}
dM = TdS + \Phi dQ + \mu dN^{3/2} 
\label{first4D}
\end{equation}
so that $\mu = \frac{\partial M}{\partial N^{3/2}}$. A simple calculation yields 
\begin{equation}
\mu =\frac{\sqrt{S} \left(7 \pi  N^{3/2} \left(1-\Phi ^2\right)-11S\right)}{36 \pi ^{3/2} N^{29/12}}
\label{mu4d}
\end{equation}
As appropriate in the grand canonical ensemble, the Gibbs free energy and the chemical potential can be obtained by expressing ${\mathcal G}$ and 
$\mu$ in terms of the temperature $T$, for fixed values of $\Phi$ and $N$. This can be obtained for example, by substituting the entropy from eq.(\ref{entf})
in the second of eq.(\ref{tempg}) or eq.(\ref{mu4d}). Alternatively, one can use the entropy as a parameter to obtain the Gibbs free energy or the chemical
potential as a function of the temperature. Although entirely equivalent, it is more illustrative to use the latter approach. 

The Hawking-Page temperature is defined via the vanishing of the Gibbs free energy of eq.(\ref{tempg}),\footnote{As we have mentioned, this happens for the 
branch $S_+$ of the entropy defined in eq.(\ref{entf}).} and the corresponding value of the entropy, $S_{HP}$ can be read off from that equation. 
Similarly, the vanishing of the chemical potential occurs at the value of $S_0$ determined by setting $\mu = 0$. These values are 
\begin{equation}
S_{HP} =\pi N^{3/2}\left(1-\Phi^2\right),~~S_{\mu=0} = \frac{7}{11}S_{HP}
\end{equation}
Using these values in the first of eq.(\ref{tempg}), we obtain
\begin{equation}
T_{HP} = \frac{\sqrt{1-\phi ^2}}{\pi N^{1/6}},~~T_{\mu=0} =\frac{8}{\sqrt{77}}T_{HP}
\end{equation}
It is thus seen that the temperature at which the chemical potential becomes zero (i.e changes sign) is always less than the Hawking-Page phase
transition temperature by a factor of $8/\sqrt{77}\sim 0.91$, and might seem to indicate that this always occurs in an unphysical region, where pure AdS 
is preferred over the black hole. 

Now we show that the situation changes qualitatively if we consider the mass, entropy, and charge densities. For the four dimensional
bulk that we consider, the dual CFT lives in a volume $V \sim L^2$ (again we will ignore some numerical pre factors that will not be important in
our analysis). The Gibbs free energy is then replaced by the Gibbs free energy density, obtained by dividing the
former by a factor of $L^2$ and does not affect the Hawking-Page phase transition temperature. But the story is different as far as the chemical potential
is concerned. This is now obtained from the first law of thermodynamics, which reads \footnote{It might be argued that we should ideally consider the
``density'' of the number of degrees of freedom. If we do use such a definition in the first law, it is straightforward to convince oneself that the 
numerical value of the chemical potential might change by an appropriate power of $N$. This does not affect the temperature at which $\mu$ changes
sign, which is what we will be interested in. We will thus continue to define $\mu$ via eq.(\ref{firstlawden}).}
\begin{equation}
d\rho = Tds + \Phi dq + \mu dN^{3/2}
\label{firstlawden}
\end{equation}
where $\rho$, $s$ and $q$ denote the mass, entropy and charge densities respectively. It is straightforward to check that in this case, 
the chemical potential (we will keep calling this by $\mu$ in order to avoid cluttering of notation) is given by 
\begin{equation}
\mu = \frac{\sqrt{s} \left(5 N^{7/6} \left(1-\Phi ^2\right)-36 s\right)}{72\pi  N^{9/4}}
\label{mu4Dden}
\end{equation}
The difference in the expressions of eq.(\ref{mu4d}) and (\ref{mu4Dden}) arises simply due to the fact that the chemical potential involves a derivative
with respect to the number of colours, and hence assumes a different value once the densities are involved (as the AdS radius $L$ is related to the number
of colours via the AdS-CFT dictionary). From eq.(\ref{mu4Dden}), we see that the chemical potential changes sign at $s = \frac{5}{36} N^{7/6}\left(1-\Phi ^2\right)$
and we find that this corresponds to a temperature $T_{\mu=0}=\frac{2}{\sqrt{5}}T_{HP}$, i.e $\sim 0.89T_{HP}$. This is different from the factor of $0.91$ that
we obtained in our previous analysis. 

In what follows, we will confine ourselves to the case where the chemical potential is defined via eq.(\ref{firstlawden}). We now consider the five dimensional
RN-AdS black hole, where the effect of considering the densities is more drastic. The procedure adopted here is similar to the one outlined for the four 
dimensional examples, and we simply record the expressions for the temperature, Gibbs free energy density and the chemical potential, in terms of the entropy 
density $s$, the electric potential $\Phi$ and the number of colours $N$. These are 
\begin{eqnarray}
T &=& \frac{N^{1/6} \left(3-4\Phi^2 + 12\times 2^{1/3}N^{-5/6}s^{2/3}\right)}{6 \times 2^{2/3} \pi s^{1/3}}\nonumber\\
g &=& \frac{N^{1/6} s^{2/3} \left(3-4\Phi^2-6 \times 2^{1/3} N^{-5/6}s^{2/3}\right)}{12\times 2^{2/3} \pi }\nonumber\\
\mu &=& \frac{s^{2/3} \left(2^{2/3} N^{5/6} \left(3-4 \Phi ^2\right)-48s^{2/3}\right)}{96\times2^{1/3} \pi  N^{8/3}}
\label{5Dquan}
\end{eqnarray}
As before, the Hawking-Page phase transition temperature $T_{HP}$ can be obtained from the zero of the Gibbs free energy density, and this can be 
compared with the temperature at which the chemical potential changes sign, $T_{\mu=0}$. The corresponding entropy densities read
\begin{equation}
s_{HP} = \frac{1}{12\sqrt{3}}N^{5/4}\left(3 - 4\Phi^2\right)^{3/2},~~s_{\mu=0} = \frac{1}{96\sqrt{3}}N^{5/4}\left(3 - 4\Phi^2\right)^{3/2}
\label{ent5D}
\end{equation}
Both of these can be seen to give the same temperature 
\begin{equation}
T_{HP} = T_{\mu=0} = \frac{3\left(3-4\Phi^2\right)}{2\sqrt{3}\pi N^{1/4}\left(3-4\Phi^2\right)^{1/2}}
\end{equation}
This might seem a little surprising, given that the entropy densities of eq.(\ref{ent5D}) differ by a factor of $1/8$. The resolution is as follows. 
Inverting the first of eqs.(\ref{5Dquan}), it can be seen that there are two solutions to the entropy as in the four dimensional example discussed before.
These are (apart from a multiplicative factor of $1/864$) :
\begin{equation}
s_{\pm} = N^{5/4} \left(3 \pi  N^{1/4} t \pm \sqrt{3} (3 \pi ^2 \sqrt{N} t^2+8\Phi ^2-6)^{1/3}\right)^3
\end{equation}
\begin{figure}[t!]
\begin{minipage}[b]{0.5\linewidth}
\centering
\includegraphics[width=2.7in,height=2.3in]{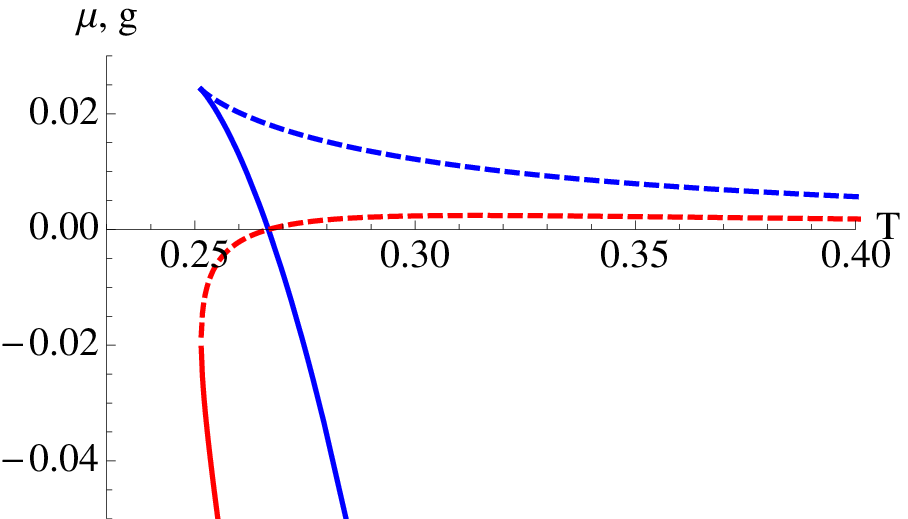}
\caption{Gibbs potential (blue, scaled by $0.5$) and chemical potential $\mu$ (red, scaled by $80$) for 5-D RN-AdS black holes as a function of the temperature, 
for the values $N=10$ and $\Phi=0.1$.}
\label{5D1}
\end{minipage}
\hspace{0.2cm}
\begin{minipage}[b]{0.5\linewidth}
\centering
\includegraphics[width=2.7in,height=2.3in]{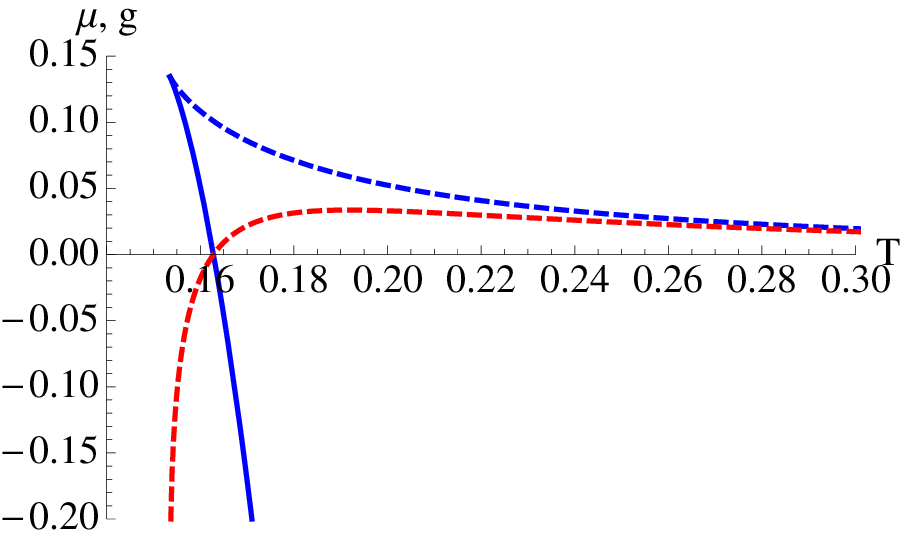}
\caption{Gibbs potential (blue, scaled by $5$) and chemical potential $\mu$ (red, scaled by $8\times 10^3$) for 5-D RN-AdS black holes as a function of temperature,
for the values $N=20$ and $\Phi=0.6$.}
\label{5D2}
\end{minipage}
\end{figure}
While the Hawking-Page phase transition always occurs on the branch corresponding to $s_+$, the chemical potential goes to zero on the branch $s_-$. 
We show this explicitly in figs.(\ref{5D1}) and (\ref{5D2}). In fig.(\ref{5D1}), we have chosen $N=10$ and $\Phi=0.1$. In this figure, the Gibbs free energy,
drawn in blue, is scaled by a factor of $0.5$, and the chemical potential drawn in red, by a factor of $80$ for better visibility (this scaling does not affect the 
sign change of these quantities, which is our focus here). The dashed and the solid lines correspond to $s=s_+$ and $s=s_-$ respectively, in their
expressions of eq.(\ref{5Dquan}). We see that both these quantities change sign at the same temperature, as alluded to above, and that whereas for 
$g$, the sign change occurs for the branch $s_+$, it occurs in the branch $s_-$ for $\mu$. In fig.(\ref{5D2}), the same quantities are plotted,
for $N=20$ and $\Phi = 0.6$. Here, $g$ is scaled by a factor of $5$ and $\mu$ by a factor of $8\times 10^3$. The same qualitative behaviour
is seen in this case as well.\footnote{We note that there is a critical value of $\Phi = \sqrt{3}/2$ beyond which $g$ and $\mu$ are always negative.}

As alluded to in the introduction, a vanishing chemical potential signals the onset of quantum effects. What we have established is that for five dimensional
RN-AdS black holes, this occurs precisely at the Hawking-Page phase transition. The same result holds after setting $Q=\Phi=0$, i.e for the five
dimensional AdS-Schwarzschild black hole as well. This result differs from the that of \cite{Dolanbose} where the chemical potential for the five dimensional
AdS-Schwarzschild black hole was defined via the thermodynamic variables, rather than their densities, and it was found that the sign change of $\mu$ occured 
in for temperatures less than $T_{HP}$.

\section{Rotating AdS Black Holes in Four and Five Dimensions}

Now we turn to the case of rotating black holes. We will first consider Kerr-AdS black holes in four dimensions. The metric is standard, and given by \cite{gibbonspope}
\begin{eqnarray}
ds^2 &=& -\frac{\Delta}{\rho^2}\left(dt-\frac{a}{\Xi}\sin^2 \theta d\phi \right)^2+\frac{\rho^2 dr^2}{\Delta}+\frac{\rho^2 d\theta^2}{\Delta_{\theta}}\nonumber\\
&+&\frac{\Delta_{\theta}\sin^2 \theta}{\rho^2}\left(a dt-\frac{r^2+a^2}{\Xi}d\phi \right)^2
\end{eqnarray}
where we have defined
\begin{eqnarray}
\Delta &=& (r^2+a^2)\left(1+\frac{r^2}{L^2} \right)-2 m r,~~\Delta_{\theta} = 1-\frac{a^2}{L^2}\cos^2\theta\nonumber\\
\rho^2 &=& r^2+a^2\cos^2\theta,~~\Xi = 1-\frac{a^2}{L^2}
\end{eqnarray}
We first record the expressions for
the relevant thermodynamic quantities, the mass $M$, entropy $S$, temperature $T$ and angular velocity $\Omega$ in terms of the rotation parameter $a$
and the horizon radius $r_h$ as\footnote{Note that the angular velocity is measured with respect to a static observer at infinity \cite{gibbonspope}.} 
\begin{eqnarray}
M &=& \frac{L^2 \left(a^2+r_h^2\right) \left(r_h^2+L^2\right)}{2 G_4\left(a^2-L^2\right)^2 r_h},~~
S = \frac{\pi  \left(a^2+r_h^2\right)}{G_4 \left(1-\frac{a^2}{L^2}\right)},\nonumber\\
T &=& \frac{a^2 \left(r_h^2-L^2\right)+r_h^2 \left(3 r_h^2+L^2\right)}{4 \pi L^2 r_h \left(a^2+r_h^2\right)},~~
\Omega = \frac{a \left(\frac{r_h^2}{L^2}+1\right)}{a^2+r_h^2}
\label{thermo}
\end{eqnarray}
We will also need the expression for the angular momentum $J$ and Gibbs free energy ${\mathcal G}= M - TS - \Omega J$. These are expressed in 
terms of the same variables $a$ and $r_h$, and read 
\begin{equation}
J = \frac{a L^2 \left(a^2+r_h^2\right) \left(r_h^2+L^2\right)}{2\left(a^2-L^2\right)^2 r_h},~~
{\mathcal G} = -\frac{\left(a^2+r_h^2\right) \left(L^2-r_h^2\right)}{4 G_4\left(a^2-L^2\right) r_h}
\label{jandg}
\end{equation}
In order to switch to physical variables, we solve for $r_h$ and $a$ in terms of $S$ and $\Omega$. The solutions read 
\begin{equation}
r_h = \frac{\sqrt{G_4} \sqrt{S} \sqrt{G_4 S -G_4 L^2 S \Omega ^2+\pi L^2}}{\sqrt{\pi  G_4 S+\pi ^2 L^2}},~~
a =\frac{G_4 L^2 S \Omega }{G_4 S+\pi  L^2}
\label{arsol}
\end{equation}
Similar expressions can obtained for $r_h$ and $a$, solved in terms of $S$ and $J$. This will be important in calculating the chemical potential
in what follows. 

As before, we use the AdS-CFT dictionary to express the thermodynamic variables in terms of the eleven dimensional Planck length and the eleven dimensional
Newton's constant. Also, we use the mass density $\rho$, the entropy density $s$ and the angular momentum density $j$ for the analysis that follows. 
It can be verified that the Hawking-Page phase transition occurs at the temperature 
\begin{equation}
T_{HP} = \frac{2-N^{1/3} \Omega ^2 +2\sqrt{1-N^{1/3} \Omega ^2}}{2 \pi N^{1/6} \left(\sqrt{1-N^{1/3} \Omega ^2}+1\right)}
\end{equation}
The temperature at which the chemical potential conjugate to the number of colours changes sign (denoted by $T_{\mu=0}$) can be straightforwardly computed, 
but the expression is lengthy and we do not show it here. We will rather present some limiting expressions which will serve to illustrate out point. 
First let us consider small values of the angular velocity $\Omega$.\footnote{We are actually considering small values of the dimensionless quantity
$\Omega l_{11}$. Since $l_{11}$ is set to unity, this fact is suppressed in what follows.} In this case, we get the following expansion for the Hawking-Page
phase transition temperature :
\begin{equation}
T_{HP} = \frac{1}{\pi N^{1/6}}-\frac{N^{1/6} \Omega ^2}{4 \pi}+{\mathcal O}\left(\Omega ^3\right)
\end{equation}
On the other hand, in this limit we obtain 
\begin{equation}
T_{\mu=0} = \frac{2}{\sqrt{5} \pi N^{1/6}}-\frac{\left(\sqrt{5}N^{1/6}\right) \Omega ^2}{9 \pi }+{\mathcal O}\left(\Omega ^3\right)
\end{equation}
The above expressions indicate that at low values of the angular velocity, $T_{\mu=0}$ is always less than $T_{HP}$, indicating that quantum 
effects become large in the region where pure AdS is preferred over the black hole. The situation however changes for large $\Omega$. Note that
for four dimensional rotating black holes physicality of the solutions demand that $\Omega < 1/L$, which translates into $\Omega < N^{-1/6}$. Hence,
we look at the region where the angular velocity is close to $N^{-1/6}$, and obtain 
\begin{eqnarray}
T_{HP} &=& \frac{1}{2 \pi N^{1/6}}+\frac{\sqrt{\alpha}}{\sqrt{2} \pi  N^{1/12}}-\frac{N^{1/12} \alpha^{3/2}}{4\sqrt{2} \pi }+
{\mathcal O}\left(\alpha^{5/2}\right)\nonumber\\
T_{\mu=0} &=& \frac{0.17}{N^{1/6}}+0.38 \alpha-0.92N^{1/6} \alpha^2+{\mathcal O}\left(\alpha^3\right)
\end{eqnarray}
where we have defined $\alpha = \left(N^{-1/6} - \Omega\right)$. With $1/(2\pi) \sim 0.16$, it is clear that close to the the maximum value of $\Omega$,
$T_{\mu=0}$ is greater than $T_{HP}$, i.e quantum effects start becoming important in a physical region, where the black hole is preferred over
a pure AdS background. This is different from the charged black hole situation. We thus see that rotation plays an important role in determining the 
onset of quantum effects in black hole phase transitions in four dimensions. 

A similar analysis holds for five dimensional Kerr-AdS black holes. The metric here is given by \cite{gibbonspope}
\begin{eqnarray}
ds^2 &=& \frac{\Delta}{\rho^2}\left(dt-\frac{a\sin^2\theta}{\Xi}d\phi \right)^2+\frac{\Delta_{\theta}\sin^2\theta}{\rho^2}\left(a dt-\frac{r^2+a^2}{\Xi}d\phi \right)^2\\&&+\left(\frac{\Delta_{\theta}r^2\cos^2\theta}{\rho^2}+\frac{\left(1+\frac{r^2}{L^2} \right)a^2 r^2 \cos^4\theta d\psi^2}{\rho^2}\right)d\psi^2+\frac{\rho^2 dr^2}{\Delta}+\frac{\rho^2 d\theta^2}{\Delta_{\theta}}\\
\end{eqnarray}
where,
\begin{eqnarray}
\Delta &=& \left(r^2+a^2 \right)\left(1+\frac{r^2}{L^2} \right)-2 m,~~\Delta_{\theta} = 1-\frac{a^2\cos^2\theta}{L^2}\nonumber\\
\rho^2 &=& r^2+a^2\cos^2\theta,~~\Xi = 1-\frac{a^2}{L^2}
\end{eqnarray}
The relevant thermodynamic quantities, i.e the mass $M$, angular momentum $J$, entropy $S$, temperature $T$ and angular velocity $\Omega$  
in terms of the rotation parameter $a$ and the horizon radius $r_h$ read :
\begin{eqnarray}
M &=& \frac{\pi  \left(3 L^2 - a^2\right) \left(a^2+r_h^2\right)\left(L^2+r_h^2\right)}{8 G_5 \left(a^2-L^2\right)^2},~~
J = \frac{\pi  a \left(a^2+r_h^2\right) \left(\frac{r_h^2}{L^2}+1\right)}{4G_5 \left(1-\frac{a^2}{L^2}\right)^2}\nonumber\\
T &=& \frac{r_h \left(a^2+2 r_h^2+L^2\right)}{2 \pi  L^2\left(a^2+r_h^2\right)},~~
\Omega = \frac{a \left(\frac{r_h^2}{L^2}+1\right)}{a^2+r_h^2},~~
S = \frac{\pi ^2 r_h \left(a^2+r_h^2\right)}{2 G_5\left(1-\frac{a^2}{L^2}\right)} 
\end{eqnarray}

In this case however, solving simultaneously for the rotation parameter $a$ and the horizon radius $r_h$ in terms of $\Omega$ and $J$ (as in eq.(\ref{arsol}) for the 
four dimensional example) is difficult. Thus we first solve for $a$ in terms of $r_h$, $\Omega$ and $L$, and feed this back in the expression for the temperature. 
This latter relation is then inverted to obtain $r_h$ in terms of $\Omega$, $T$ and $L$. This information is then used to compute the Gibbs free energy, which in
terms of $r_h$ and $a$ has a simple form
\begin{equation}
{\mathcal G} = \frac{\pi  \left(a^2+r_h^2\right) \left(r_h^2-L^2\right)}{8 G_5\left(a^2-L^2\right)}
\end{equation}
Then using the AdS-CFT dictionary, to set $L \sim N^{1/4}$, we obtain the Gibbs free energy in terms of $\Omega$, $T$ and $N$. Finally, the chemical
potential is obtained via $\mu=\left(\partial g/\partial N^2\right)_{T,\Omega}$ where $g$ is the Gibbs free energy density 
(note that the number of degrees of freedom scales as $N^2$ in five dimensions). 
\begin{figure}[t!]
\begin{minipage}[b]{0.5\linewidth}
\centering
\includegraphics[width=2.7in,height=2.3in]{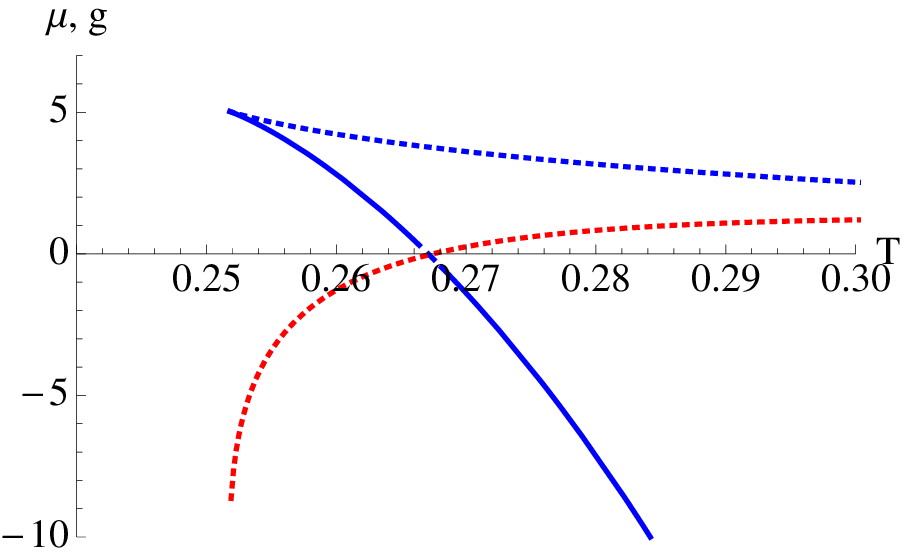}
\caption{Gibbs potential density (blue, scaled by $10^2$) and chemical potential $\mu$ (red, scaled by $4\times10^4$) for 5-D Kerr-AdS black holes 
as a function of the temperature, for the values $N=10$ and $\Omega=0.1$.}
\label{5Drot1}
\end{minipage}
\hspace{0.2cm}
\begin{minipage}[b]{0.5\linewidth}
\centering
\includegraphics[width=2.7in,height=2.3in]{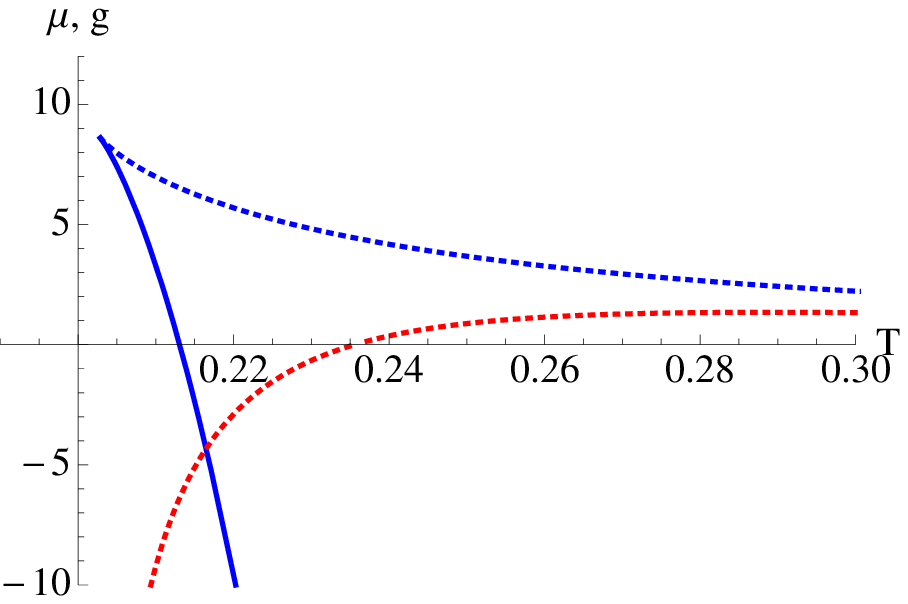}
\caption{Gibbs potential density (blue, scaled by $10^2$) and chemical potential $\mu$ (red, scaled by $4\times10^4$) for 5-D Kerr-AdS black holes 
as a function of temperature, for the values $N=10$ and $\Phi=0.52$.}
\label{5Drot2}
\end{minipage}
\end{figure}
The expressions are rather tedious, and we resort to a graphical analysis. In figs.(\ref{5Drot1}) and (\ref{5Drot2}), we show the variation of the Gibbs free energy
density (blue) and the chemical potential (red) as a function of the temperature for two different values $\Omega = 0.1$ and $0.52$ respectively. In both
cases, we have taken $N=10$. For small values of $\Omega$, the chemical potential becomes positive very close to the Hawking-Page temperature $T_{HP}$,
as expected from the AdS-Schwarzschild analysis. However, as we increase $\Omega$ towards its maximum value $(=N^{-1/4} \sim 0.56)$, $T_{\mu=0}$ is seen 
to be greater than $T_{HP}$. In these figures, the dotted and solid lines correspond to the higher and lower entropy branches, and as before, the $T_{HP}$ 
($T_{\mu=0}$) occurs in the higher (lower) entropy branch, respectively. 

\section{AdS Black Holes in Einstein-Gauss-Bonnet Gravity}

We finally come to the case of the Gauss Bonnet black holes in five dimensions. 
The Einstein-Maxwell Gauss-Bonnet action in five dimensions with a cosmological constant $\Lambda$ is given by \cite{Cai2002}
\begin{equation}
S=\frac1{16\pi G_5}\int d^5x \sqrt{-g}[R-2\Lambda+\frac{\alpha}{2} (R_{\mu\nu\gamma\delta}R^{\mu\nu\gamma\delta}-
4R_{\mu\nu}R^{\mu\nu}+R^2)-4\pi F_{\mu\nu}F^{\mu\nu}]
\end{equation}
where $G_5$ is Newton's constant in $d=5$ and $F_{\mu\nu}$ is the field strength tensor and $\alpha$ is the Gauss-Bonnet parameter.
\begin{figure}[t!]
\begin{minipage}[b]{0.5\linewidth}
\centering
\includegraphics[width=2.7in,height=2.3in]{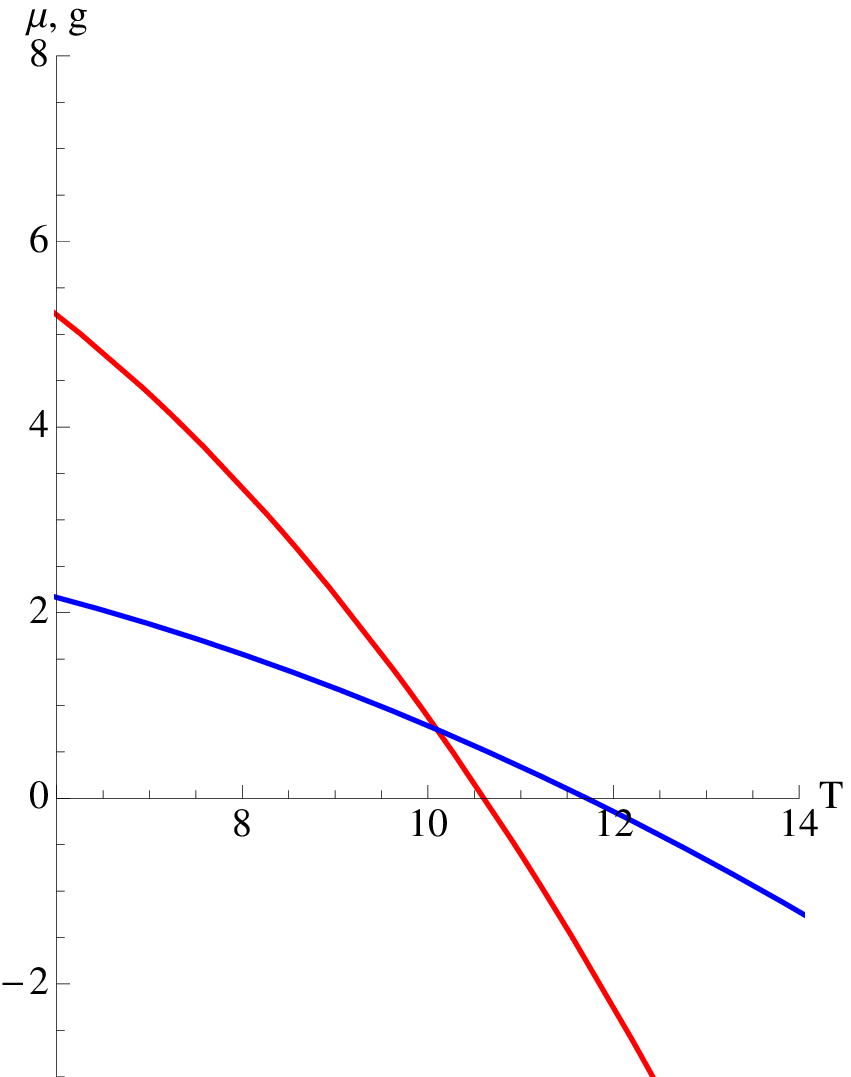}
\caption{Gibbs potential density (blue, scaled by $10^2$) and chemical potential $\mu$ (red, scaled by $10^3$) for 5-D AdS-Gauss Bonnet black holes as a function of the 
temperature (scaled by $10^2$), for the values $N=1$, $\alpha=0.5$.}
\label{GB1}
\end{minipage}
\hspace{0.2cm}
\begin{minipage}[b]{0.5\linewidth}
\centering
\includegraphics[width=2.7in,height=2.3in]{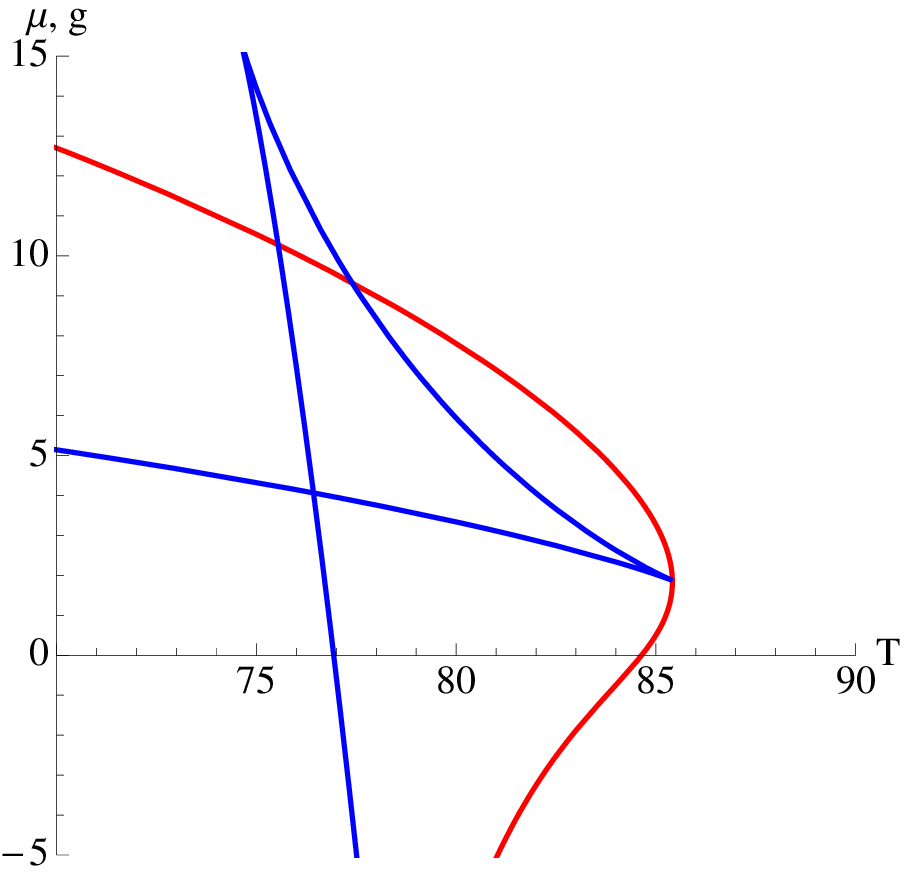}
\caption{Gibbs potential density (blue, scaled by $10$) and chemical potential $\mu$ (red, scaled by $10^8$) for 5-D AdS-Gauss Bonnet black holes as a function of the
temperature (scaled by $10^3$) for the values $N=1000$ and $\alpha=0.5$.}
\label{GB2}
\end{minipage}
\end{figure}
The action has a well known solution given by the metric
\begin{equation}
ds^2=-f(r)dt^2+f^{-1}(r)dr^2+r^2d\Omega_3^2,
\end{equation}
where $d\Omega_3$ represents the line element of the 3-sphere and
\begin{equation*}
f(r)=1+\frac{r^2}{2 \alpha}\left (1-\sqrt{1+\frac{32  \alpha M G_5}{3 \pi r^{4}}-\frac{16 \alpha Q^2 G_5^2}{3 \pi^2 r^{6}}-\frac{4 \alpha}{L^2}} \right )
\end{equation*}
$M$ is the thermodynamic mass and $Q$ is the electric charge of the five-dimensional Gauss-Bonnet black hole with spherical event horizon topology. 
As before, we will denote $r_h$ as the radius of the event horizon. In terms of $r_h$ and $Q$, the thermodynamic mass $M$, electric potential $\Phi$,
temperature $T$ and entropy $S$ are given as
\begin{eqnarray}
M &=& \frac{3 \pi  r_h^2}{8G_5} \left(1 + \frac{\alpha }{r_h^2} +\frac{r_h^2}{L^2} \right)+\frac{G_5 Q^2}{2 \pi  r_h^2},~~
\Phi = \frac{G_5 Q}{\pi  r_h^2}\nonumber\\
T &=& \frac{3 \pi ^2 r_h^4 \left(2 r_h^2+L^2\right)-4 G_5^2 L^2 Q^2}{6 \pi ^3 L^2 r_h^3 \left(r_h^2 + 2 \alpha\right)},~~
S = \frac{\pi^2 r_h^3}{2 G_5} \left(1 + \frac{6 \alpha}{r_h^2}\right),~~
\end{eqnarray}
The Gibbs free energy density is, 
\begin{equation}
g = -\frac{3 \pi ^2 r_h^2 \left(-L^2 \left(6 \alpha^2-3 \alpha  r_h^2+r_h^4\right)+18 \alpha  
r_h^4+r_h^6\right)+ 4 G_5^2 L^2 Q^2 \left(r_h^2-6 \alpha \right)}{24 \pi  G_5 L^2 r_h^2 \left(2\alpha +r_h^2\right)}
\end{equation}
We will focus on the simple case $Q=0$. In this case, one finds a subtle interplay between the two parameters of the theory, namely the Gauss-Bonnet coupling
$\alpha$ and the number of colours $N$. We first fix $\alpha$ to a small value, say $\alpha = 0.01$. In this case,  one observes a swallow-tail behaviour for the
Gibbs free energy density starting from $N=1$.\footnote{The swallow-tail is always in the region of positive Gibbs free energy density and hence possibly not
particularly interesting.} However, with $\alpha = 0.1$, this swallow tail behaviour sets in for a larger value of $N \sim 15$. 
Conversely, for a fixed $N$, the swallow-tail behaviour for a given value of $\alpha$ disappears when one increases $\alpha$. 

To discuss the nature of the chemical potential vis a vis the Gibbs free energy density, we fix $\alpha=0.5$. In figs.(\ref{GB1}) and (\ref{GB2}), we show the 
behaviour of the Gibbs free energy (blue) and the chemical potential (red) for $N=1$ and $N=1000$ respectively. In figs.(\ref{GB1}) and (\ref{GB2}), the 
Gibbs free energy density has been scaled by $10^2$ and $10$, the chemical potential by $10^3$ and $10^8$ and the temperature by $10^2$ and $10^3$
respectively, for better visibility. We see that for this value of $\alpha$, as one increases the number of colours, the chemical potential changes sign at
a temperature greater than $T_{HP}$. A qualitatively similar analysis can be straightforwardly done for non-zero charge. We will however not present
the details here.  

\section{Discussions and Conclusions}

In this paper, we have considered extended phase space thermodynamics of a class of charged and rotating AdS black holes in four and five
dimensions, and the AdS-Gauss-Bonnet black hole in five dimensions. The analysis was done in the grand canonical ensemble. 
We defined the chemical potential dual to the number of colours of the boundary
gauge theory via densities of standard thermodynamic variables. Our main conclusion here is that for five dimensional AdS-Schwarzschild and RN-AdS
black holes, this chemical potential changes sign precisely at the location of the Hawking-Page phase transition. This signals the onset of quantum
effects, since a vanishing chemical potential conventionally signals non-conservation of particle number. This is physically reasonable, and might
point to important physics at the Hawking-Page transition, which, in conventional thermodynamics, is dual to a confinement-deconfinement transition 
of the boundary gauge theory. 

For rotating black holes in four and five dimensions, our analysis shows that for a sufficiently large value of the angular frequency (close to its maximum 
value), the chemical potential changes sign in a stable black hole region, i.e above the Hawking-Page temperature. The precise angular frequency where
the potential changes sign at the black hole phase transition should be an important point to revisit. We further analysed non-rotating Gauss-Bonnet-AdS black holes
in five dimensions and saw features of the chemical potential that are similar to Kerr-AdS black holes. 

Understanding the nature of the quantum effects due to a vanishing chemical potential, from the boundary field theory perspective, should be an important 
issue for future research. 

\begin{center}
{\bf Acknowledgements}
\end{center}
The work of RM is supported by the Department of Science and Technology, Govt. of India, by the grant IFA12-PH-34.

\end{document}